\documentclass[english]{article}
\usepackage[T1]{fontenc}
\usepackage[latin1]{inputenc}
\usepackage{a4wide}
\usepackage{amsmath}
\usepackage{setspace}
\makeatletter
 \newcommand{\lyxaddress}[1]{
   \par {\raggedright #1 
   \vspace{1.4em}
   \noindent\par}
 }

\usepackage{graphicx}
\usepackage{dcolumn}
\usepackage{bm}

\usepackage{babel}
\makeatother
\begin{document}

\title{Diode Effect in Asymmetric Double Tunnel Barriers with Single Metal
Nanoclusters}

\author{A. Iovan, D. B. Haviland, and V. Korenivski}

\maketitle

\lyxaddress{Nanostructure Physics, Royal Institute of Technology, SE-10691 Stockholm,
Sweden }

\begin{abstract}
Asymmetric double tunnel barriers with the center electrode being
a metal cluster in the quantum regime are studied. The zero dimensionality
of the clusters used and the associated quantized energy spectra are
manifest in well-defined steps in the current voltage characteristic
(IVC). Record high current rectification ratios of $\sim10^{4}$ for
tunneling through such clusters are demonstrated at room temperature.
We are able to account for all of the experimentally observed features
by modeling our double barrier structures using a combination of discrete
states and charging effects for tunneling through quantum dots. \newpage

\end{abstract}
Nanometer scale electronic devices are of great interest from the
fundamental physics point of view as systems of low dimensionality,
and of technological significance since they provide the basis for
a large spectrum of new applications in microelectronics. For example,
spintronic devices \cite{Maekava}, where the conductance is a function
of the magnetic state of the device, are used in magnetic random access
memory (MRAM) designs or magnetic field sensors \cite{Parkin}. A
high relative change in current when the bias voltage is reversed
(known as the current rectification ratio, RR) is a particularly interesting
property for, e. g., realizing the select function in magnetic memory
cell design \cite{Parkin}. Current rectification has been demonstrated
for transport through quantum levels in molecules \cite{Datta,Zhao}
and semiconductor nanoparticles \cite{Schmidt,Bakkers}. However,
for MRAM it is highly desirable to integrate the rectifying device
into the metal/insulator based magnetic tunnel junction. The proposed
approach to achieve the diode function in such metal/insulator based
structures is to embed a thin metallic layer or a nanocluster into
an asymmetric double tunnel barrier \cite{Chshiev,Bonet,kanjouri},
where it is essential that the cluster is in the quantum regime with
the energy spectrum governed by dimensional or charge quantization
\cite{Thinkham,Wilkins,Ralph,Yakushiji,Graf}. Such an approach has
been used to obtain a RR of 10-100 at room temperature in relatively
large junctions \cite{Tiusan,Iovan}. The size of these junctions
($10\times10\ \mu\text{m}^{2}$) most likely resulted in tunneling
through multiple particles of different dimensions, conducting in
parallel, and thereby averaging away discrete features in the transport
characteristics . In this work we have achieved tunneling through
a single metallic nanocluster, demonstrating record RR's up to $\sim10^{4}$
at room temperature.

We use a scanning tunneling microscope (STM) based technique \cite{Wilkins}
with three layer samples. The top layer of the samples consists of
nano-particles ($\sim$ 1 nm in size) insulated by a thin insulating
layer from the bottom electrode (see Fig.1a). The width of the top
particle-to-tip tunnel barrier, can be varied over a broad range with
the STM, by varying the tip-to-surface separation. Thus, a double
tunnel junction can be formed with one fixed and one variable transparency
barrier enclosing a single metallic nanodot. In measuring the samples
we aimed to achieve high barrier asymmetry, since the theory \cite{Chshiev,Bonet}
predicts maximum RR in this regime.

The three layer samples were prepared as follows: bottom electrodes
of Co, a few tens of nanometers in thickness, were deposited by electron
beam evaporation and subsequently covered by a thin (1-2 nm) layer
of Al-O. On top of of the insulator, sub-percolation, thin layers
of Co ($\sim1$ nm) were evaporated. These Co nanoclusters, somewhat
larger in size than 1 nm, were naturally oxidized in order to further
reduce in size the metallic core. A matrix of $\sim1$ nm Co clusters
separated from the bottom electrode by a fixed transparency tunnel
barrier was thus produced. In what follows, a negative voltage corresponds
to a flow of electrons from the sample to the STM tip. All measurements
were done at room temperature in zero magnetic field. 

The physical model of the structure is illustrated in Fig. 1b. It
consists of two tunnel barriers enclosing a quantum dot, with the
tunnel barriers characterized by their transparency and capacitance.
For the Al-O barrier the transparency $\gamma_{1}$ and capacitance
$C_{1}$ are fixed. The second barrier has respectively $\gamma_{2}$
and $C_{2}$, which can be varied by varying the tip-to-particle distance.
The particle must be comparable in size to the Fermi wavelength in
order to exhibit the desired quantum size effects \cite{Thinkham,Himpsel},
which then dominate the transport through the structure. 

The surface of a sample is scanned in order to determine a suitable
location for the IVC. If the particular junction has the desired asymmetric
IVC the measurement is repeated and checked for reproducibility. A
large number of such junctions have been made and measured. Many measurements
revealed a tunneling characteristic with negligible current asymmetry,
due to either tunneling through a single barrier or a double barrier
with the center island not exhibiting quantum effects. 

A number of measurements exhibited very well defined steps in the
IVC with very high RR, such as that is illustrated in Fig. 2a. A staircase
in current is observed for positive bias which is due to dimensional
or charge quantization, or a combination of these two effects \cite{Bonet}.
The Coulomb blockade energy $E_{c}=e^{2}/2C_{\Sigma}$ is approximately
0.7 eV for this double-junction, corresponding to a particle size
of 1-3 nm. This estimate is based on the approximation of spherical
particles where, $C_{\Sigma}=8\epsilon R$, with R is the radius of
the particle and $\epsilon_{r}=8-10$ the dielectric constant for
Al-O. As seen in Fig 2a, the sharp jumps in the current (peaks in
transmission) are separated by 60-200 mV which gives the scale of
the energy and level spacing on the cluster. The shape of the IVC
is a sensitive function of the relative transparencies of the two
barriers \cite{Schmidt,Su}. Charge on the center island is accumulated
differently for positive and negative bias voltages if the barriers
have significantly different transparencies. In the case where the
transparency for the incoming electrons is high and that for the outgoing
electron is low, the first tunneling event results in a long lived
electron state on the island, which blocks subsequent single electron
transport. In the opposite case of low incoming and high outgoing
transparency, the charge accumulation on the island is negligible.
Thus barrier asymmetry explains our observations of current jumps
for positive bias polarity only, which reflects the situation with
a rapid entrance of electrons onto, and slow escape from the island.This
asymmetry also results in a large RR (the modulus of the ratio of
currents for the two bias directions) for the IVC of Fig. 2a, shown
in Fig. 2b as a function of bias voltage. The RR reaches $\sim10^{4}$,
which is the highest RR value reported to date for tunneling through
metallic nano-particles.

To further to interpret the measurements we modeled the transport
taking in to account both the quantization of the energy levels on
the island and the charging effects in the structure. Incorporation
of the dimensional quantization effects is necessary since the theory
of the Coulomb Blockade \cite{Ferry} for a system with two highly
asymmetric barriers predicts no asymmetry in the IVC (for island potential
of $e/2$), with the Coulomb staircase identical for both bias directions.
The resonant tunneling model without charging effects \cite{Chshiev}
can be used to explain asymmetric IVC's for a double junction with
asymmetric barriers. However, in our case of the nanometer size particles
the charging energy is expected to be significant and must be included
into the considerations. We use the model and simulation code of Bonnet
et al. \cite{Bonet} which combines both charging effects and those
due to discrete states. Fig. 3a shows a simulated highly asymmetric
IVC with discrete steps, which closely resembles our experimental
data (see Fig. 2a). Our junctions meet the following requirements
for validity of the model: $k_{B}T$ at the measurement temperature
(300 K) is smaller than the energy level spacing, the level spacing
is smaller than the Coulomb gap, and the barrier resistance is higher
than the resistance quantum $h/e^{2}$. The calculated IVC of Fig
3 assumes that the cluster has 7 equidistant levels with the spacing
of 200 mV, starting at 300 meV above the Fermi level of the outer
electrodes. We use the following parameters for the double barrier:
$\gamma_{1}/\gamma_{2}=10$ with $\gamma_{1}=110$ MHz and $C_{2}/C_{1}=4$.
From the calculated IVC we obtain the RR as a function of the bias
voltage, which is shown in Fig. 3b. Its functional form and the magnitude
at the maximum is in good agreement with the experimental data shown
in Fig. 2b. The good agreement between the theoretical and experimental
data strongly suggests that we indeed observe quantized transport
through a single nano-cluster. In the simulation we assume that the
gate charge (or offset charge) is zero. Random offset charge due to
possible defects in the insulator and/or electrostatic coupling from
neighboring nanoparticles, can affect the potential on the center
island and generally with gate charge. However, we argue that the
offset charge in such nanometer small grains should be a small fraction
of $e$ since the very small $C_{\Sigma}$, and therefore large $E_{C}=e²/2C_{\Sigma}$,
lead to a weak coupling of the grain to random offseting charges that
can be present in its vicinity.

The distance between the energy levels is directly related to the
particle size, which is illustrated by the following consideration
\cite{Halperin}. The conduction electron spectrum of a bulk metal
is continuous. For small metallic particles, however, the spectrum
is discrete. In the free electron model, the Fermi energy depends
only on the electron density and consequently is not a function of
the particle size. However, the number of conduction electrons is
small and therefore the number of the filled electronic levels is
small. Since all states are filled up to $E_{F}$, the level spacing
increases with decreasing the size of the particle approximately as
$\delta\approx E_{F}/N$, where $N$ is the number of conduction electrons
in the particle. With $E_{F}\sim10$ eV and the electron density of
$n\sim10^{29}$ m$^{-3}$, $N\sim100$ for a $\sim1$ nm particle
and the energy level spacing is $\delta\sim100$ meV, coinciding well
with the values used in the modelling above. 

In conclusion, using a multilayer sample with an STM technique, we
have achieved transport through discrete electronic states of metallic
nanoparticles, exhibiting quantum effects at room temperature. We
observe record high current rectification ratios of $\sim10^{4}$
and jumps in current correspond to tunneling through discrete states.
Theoretical modelling confirms that the origin of the observed behavior
in the double tunnel junctions is dimensional and charge quantization
in the center electrode combined with the tunnel barriers of different
transparency. This work experimentally demonstrates the viability
of the concept of a double junction for obtaining the diode functionality
in a metal/oxide structure. This functionality should be important
for such technologies as MRAM where the same metal/oxide stack can
be also used to perform cell select.\\
\\
 \indent This work is supported by the Swedish Foundation for Strategic
Research, the Royal Swedish Academy of Sciences and the K. A. Wallenberg
Foundation.

Captions:

Figure 1 (a) Schematic of the measurement configuration and (b) profile
of the barriers.

Figure 2 (a) IVC measured of a double tunnel junction exhibiting pronounced
current quantization and (b) RR as a function of bias voltage.

Figure 3 (a) IVC simulated of a double tunnel junction exhibiting
current quantization and (b) calculated RR as a function of bias voltage. \newpage


\newpage

\begin{thebibliography}{10}
\bibitem{Parkin}S. S. P. Parkin et al, J. Appl. Phys 85, 5828 (1999). 
\bibitem{Maekava}Sadamichi Maekawa, Teruya Shinjo, \textit{Spin Dependent Transport
in Magnetic Nanostructures}, Taylor and Francis, London (2002). 
\bibitem{Parkin}Parkin, S. S. P. et al. Proc. IEEE \textbf{91}, 661-680 (2003). 
\bibitem{Datta}Supriyo Datta, Weidong Tian, Seunghun Hong, R. Reifenberger, Jason
I. Henderson, and Clifford P.Kubiak, Phys. Rev. Lett. \textbf{79},
2530 (1997). 
\bibitem{Zhao}Jin Zhao, Changgan Zeng, Xin Cheng, Kedong Wang, Guanwu Wang, Jinlong
Yang, J. G. Hou, and Qingshi Zhu, Phys. Rev. Lett. \textbf{95}, 045502
(2005). 
\bibitem{Schmidt}T. Schmidt, R. J. Haug, K. v. Klitzing, A. Förster and H. Lüth, Phys.
Rev. B \textbf{55}, 2230 (1997). 
\bibitem{Bakkers}Bakkers Erik P. A. M. Bakkers and Daniël Vanmaekelbergh, Phys. Rev.
B \textbf{62}, 7743 (2000). 
\bibitem{Thinkham}M. Tinkham, D. Davidovic, D.C. Ralph, and C.T. Black, Journal of Low
Temperature Physics, \textbf{118}, 271 (2000). 
\bibitem{Wilkins}R. Wilkins, E. Ben-Jacob, and R. C. Jaklevic, Phys. Rev. Lett. \textbf{63},
801 (1989). 
\bibitem{Ralph}D. C. Ralph, C. T. Black, and M. Tinkham, Phys. Rev. Lett. \textbf{74},
3241 (1995). 
\bibitem{Yakushiji}Kay Yakushiji, Franck Ernult, Hiroshi Imamura, Kazutaka Yamane, Seiji
Mitani, Koki Takanashi, Saburo Takahashi, Sadamichi Maekawa, Hiroyasu
Fujimori, Nature Materials \textbf{4}, 57-61 (2005). 
\bibitem{Graf}Graf, J. Vancea and H. Hoffmann, Appl. Phys. Lett. \textbf{80}, 1264
(2002). 
\bibitem{Tiusan}C. Tiusan, M. Chshiev, A. Iovan, V. da Costa, D. Stoeffler, T. Dimopoulos,
K. Ounadjela, Appl. Phys. Lett. \textbf{79}, 4231 (2001). 
\bibitem{Iovan}A. Iovan, \textit{thesis}, IPCMS, Strasbourg (september 2004). 
\bibitem{Chshiev}M. Chshiev, D. Stoeffler, A. Vedyayev and K. Ounadjela, Europhys.
Lett. \textbf{58}, 257 - 263 (2002). 
\bibitem{Bonet}Edgar Bonet, Mandar M. Deshmukh, D. C. Ralph, Phys. Rev. B \textbf{65},
045317 (2002). 
\bibitem{kanjouri}F. Kanjouri, N. Ryzhanova, N. Strelkov, A. Vedyayev, and B. Dieny,
J. Appl. Phys. \textbf{98}, 083901 (2005).
\bibitem{Su}Su Bo Su, V. J. Goldman, J. E. Cunningham, Science \textbf{255}, 313-315
(1992). 
\bibitem{Ferry}David K. Ferry, Stephen M. Goodnick, \textit{Transport in Nanostructures},
Cambridge University Press (1997). 
\bibitem{Halperin}W.P. Halperin, Reviews of Modern Physics, \textbf{58}, 533 (1986). \newpage

\end{thebibliography}
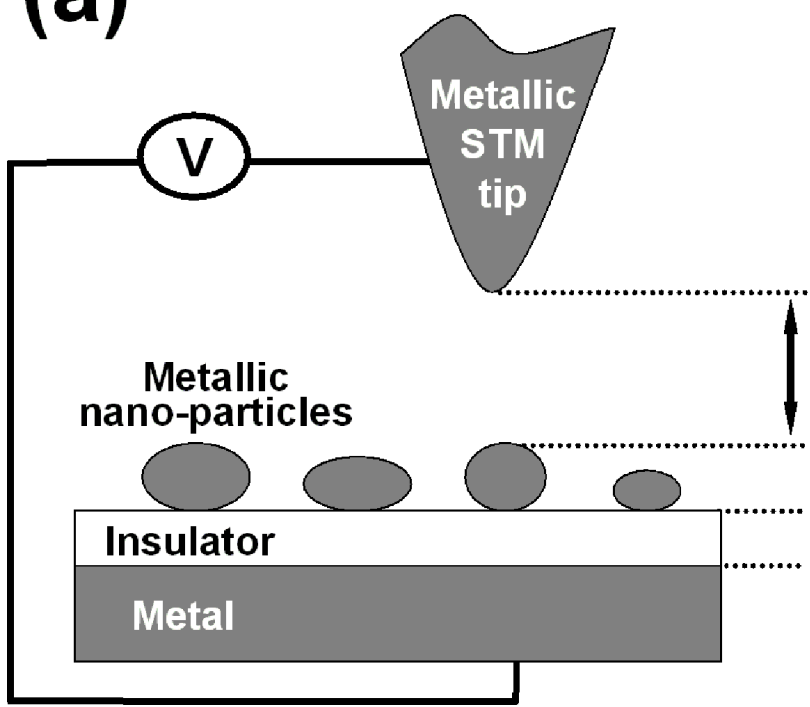
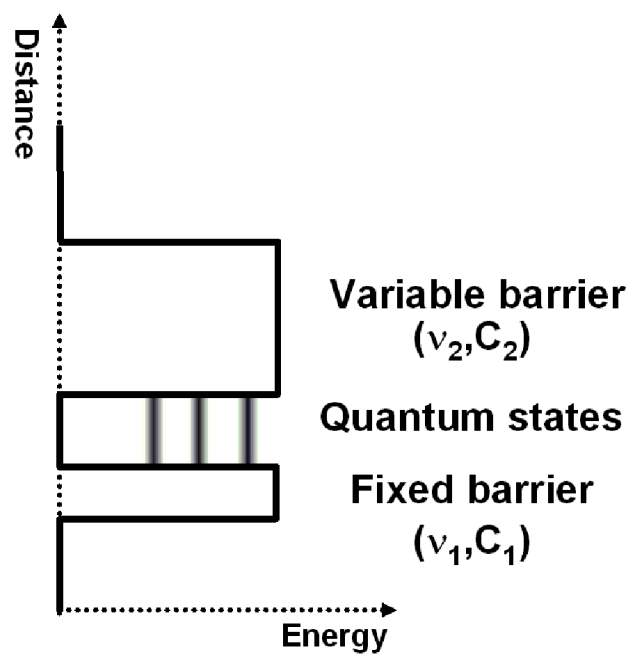
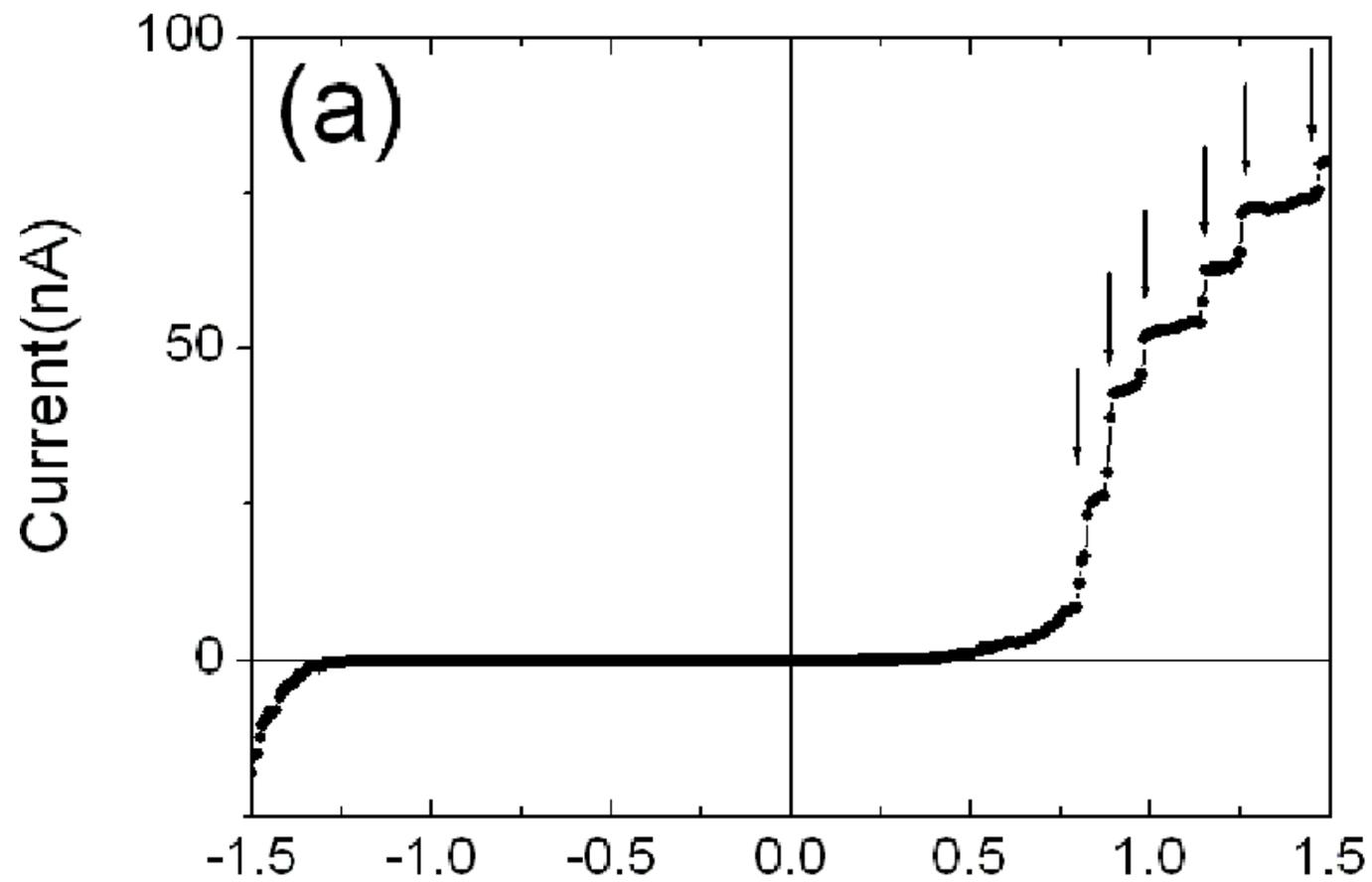
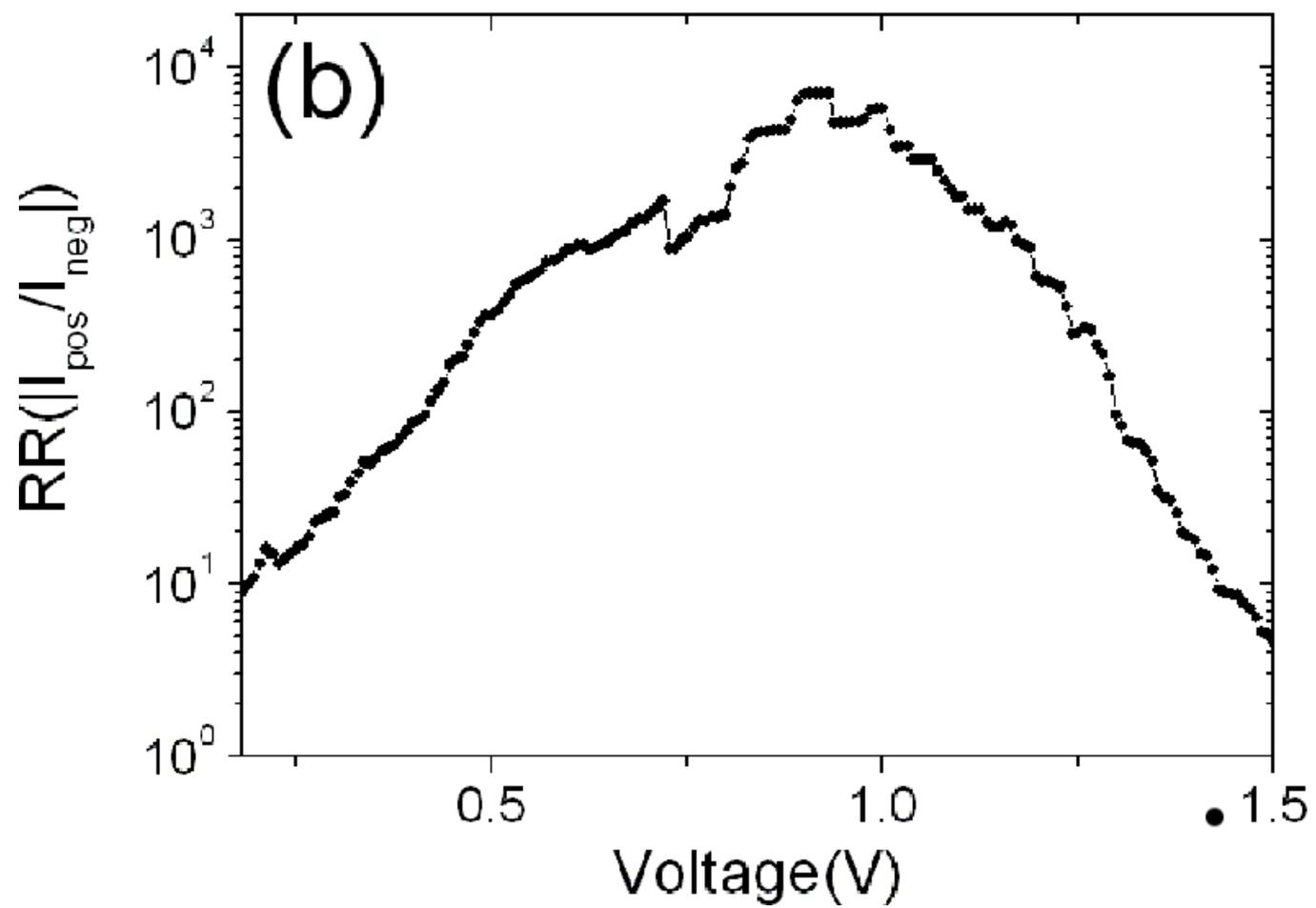
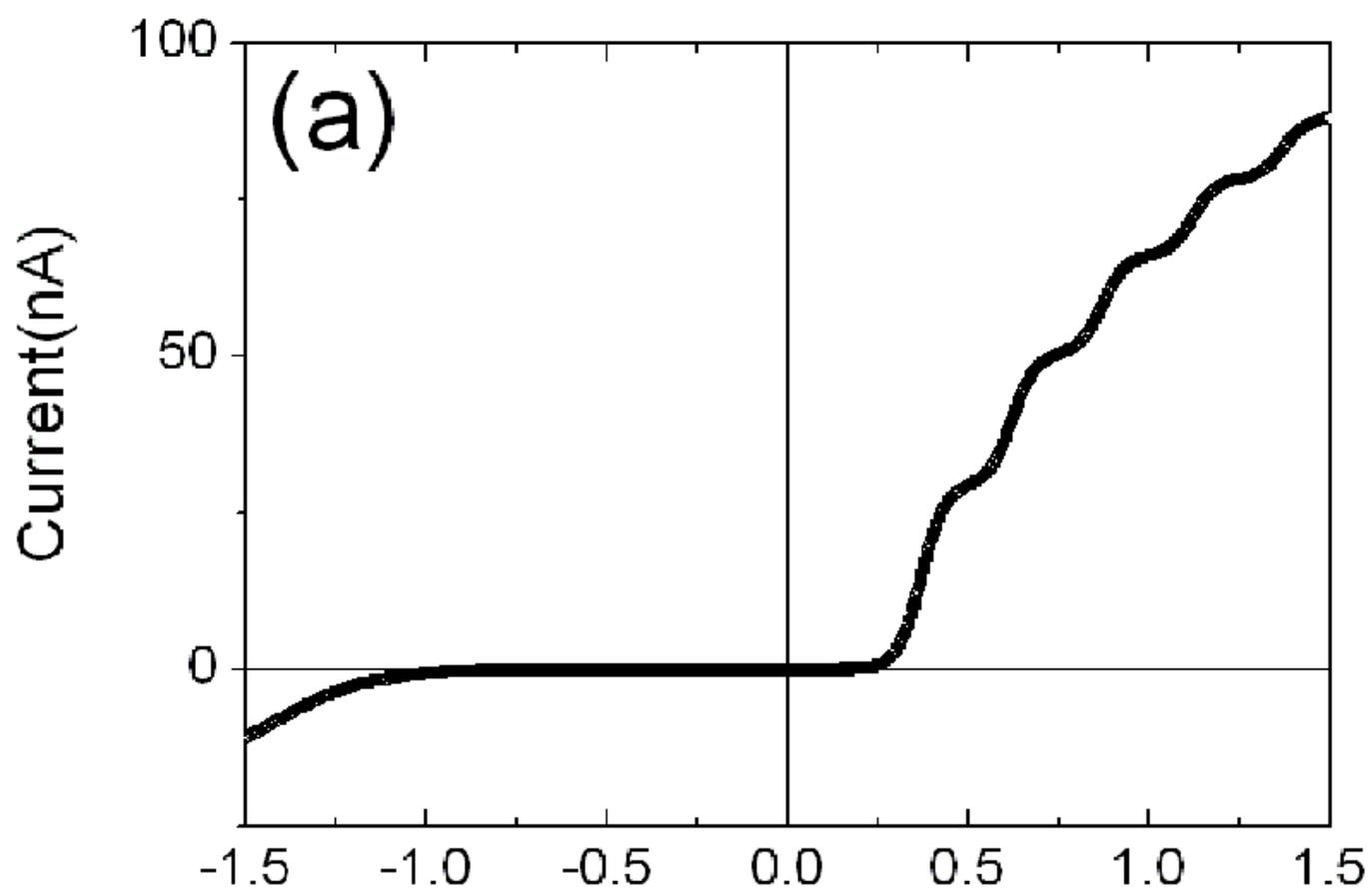
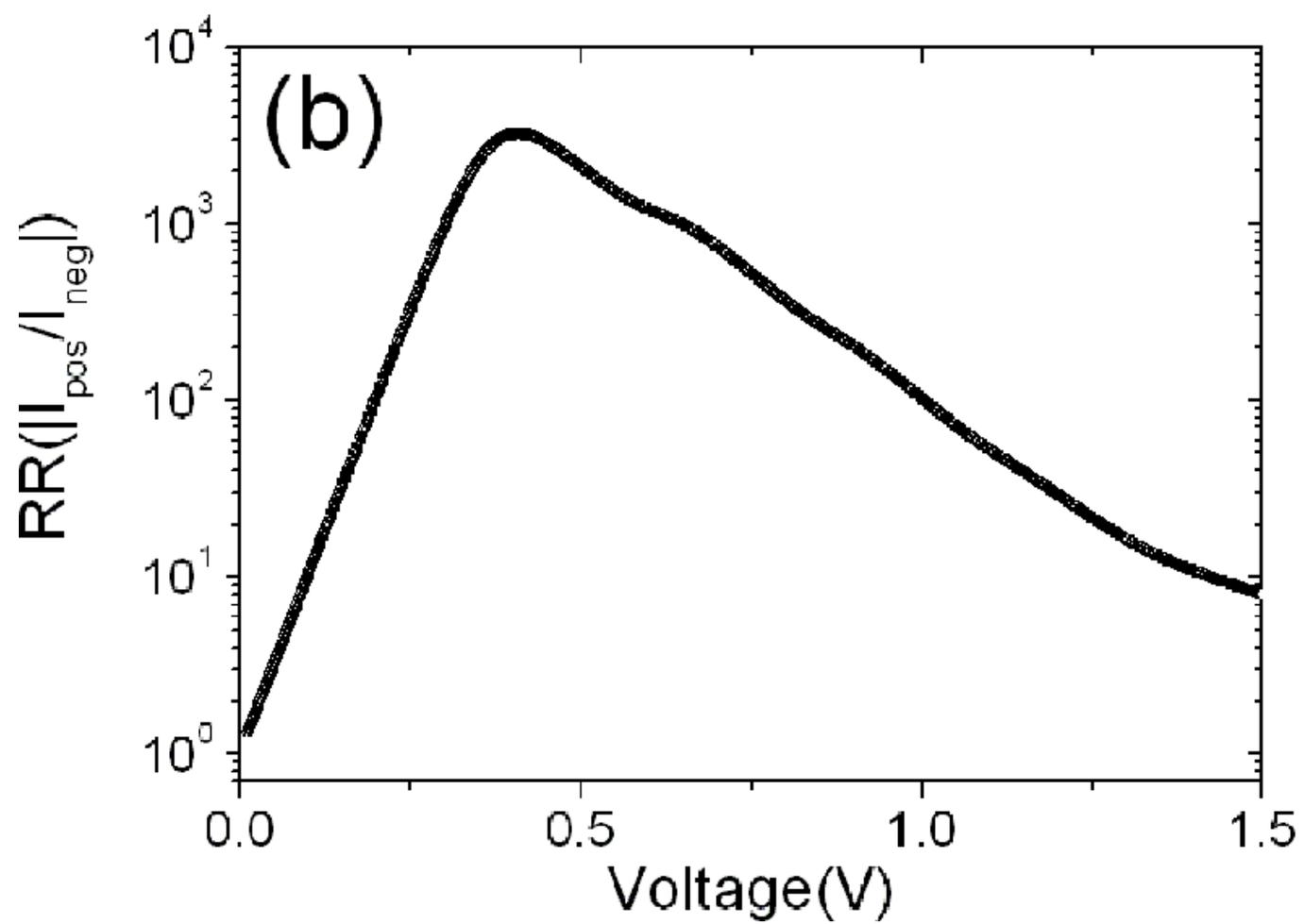
\end{document}